\documentstyle[11pt,iau185,twoside,epsf]{article}

\begin{document}
\title{Periodic and Nonperiodic Phenomena in AGB Stars\altaffilmark{1}}
\author{T. Lebzelter}
\affil{Institut f. Astronomie, University of Vienna, Austria}
\author{K.H. Hinkle}
\affil{National Optical Astronomy Observatories\altaffilmark{2}, Tucson, USA}
\altaffiltext{1}{Based on observations obtained at Kitt Peak National
Observatory}
\altaffiltext{2}{operated by the Association of Universities for Research
in Astronomy, Inc. under cooperative agreement with the National Science
Foundation}

\begin{abstract}
We discuss velocity variations in different layers of an AGB star's atmosphere.
Periodic and nonperiodic changes are observed. From a large sample of data
we derive mean velocity curves for miras and semiregular variables.
\end{abstract}

 
\section{Introduction}
The structure of the atmospheres of Asymptotic Giant Branch (AGB) stars is
dominated by stellar pulsations and therefore varies strongly during the light
cycle.  High resolution infrared spectroscopy has turned out to be a very
powerful tool to investigate the dynamics of AGB star atmospheres caused by the
stellar pulsation (Hinkle et al., 1982).  The rich
vibrational-rotational molecular spectrum of these stars offers a wide variety
of lines of different excitation originating in different layers of the
atmosphere.  These lines can be used to monitor the dynamics at different depths
in the stellar atmosphere.  However, so far only a small sample of AGB variables
and only a small number of molecular species have been investigated in detail
(e.g. Hinkle et al., 1984).

In this paper we present a summary of our recent work on the dynamics of AGB
star atmospheres obtained by means of time series high resolution infrared
spectra of a large sample of AGB stars. We extended the existing data material
significantly, investigating the behavior of a large variety of molecular lines
in AGB variables of different types.

\section{Observations}
Time series of high resolution spectra of several parts of the 1.5 to 4\,$\mu$m
region have been obtained at Kitt Peak National Observatory with the FTS, the
NICMASS- and the PHOENIX-spectrograph.  Spectral resolution was typically 40000
or better. Line positions have been measured for a selection of atomic and
molecular species. The derived velocities have an accuracy of at least 0.3
kms$^{-1}$. Several lines exhibit at some phases multiple absorption and/or
emission components. Each of these components has been measured.  Their reality
has been checked using averaged line profiles (Lebzelter, Hinkle, \& Aringer
2001).

\section{Results}

\begin{figure}
\begin{center}
\plottwo{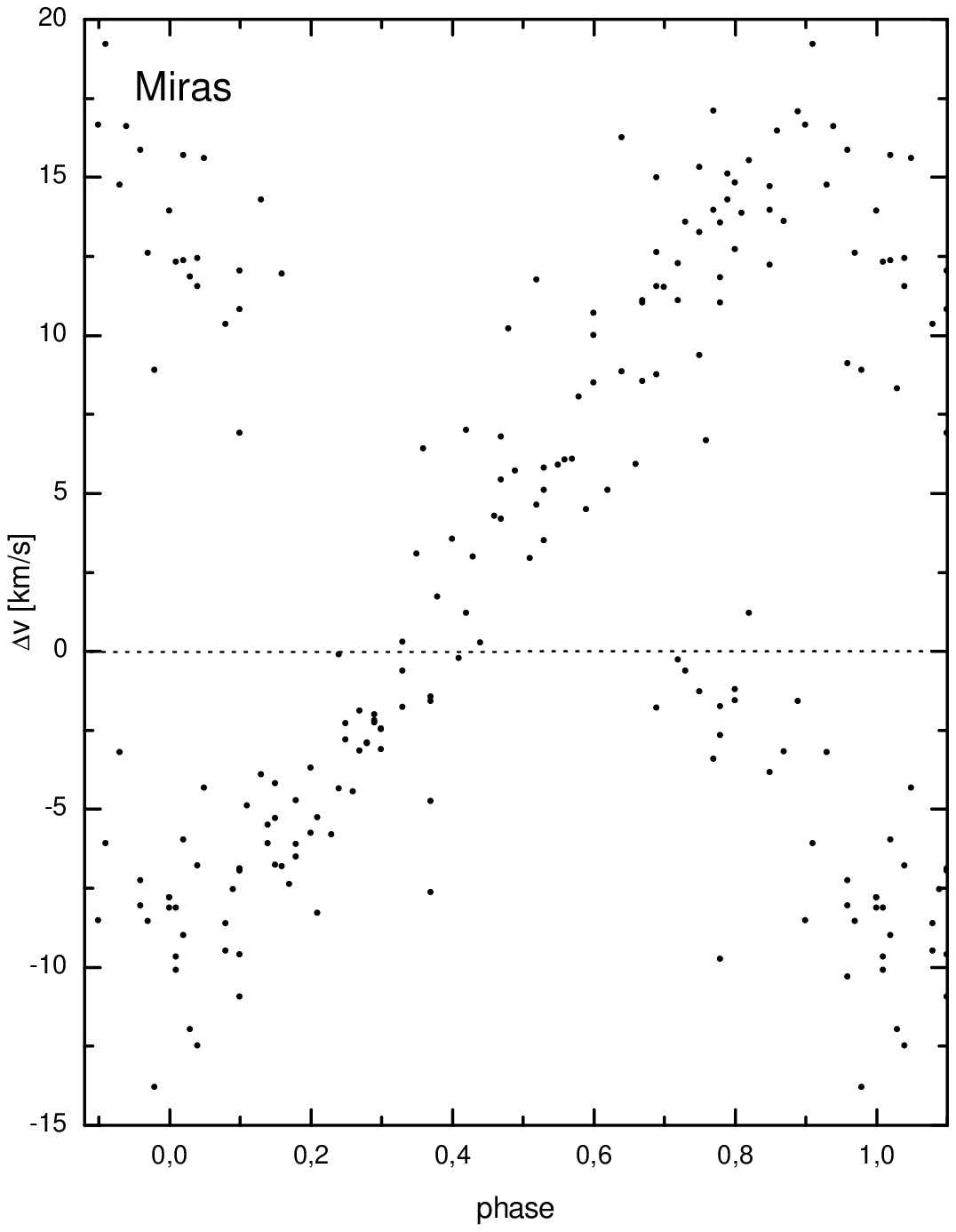}{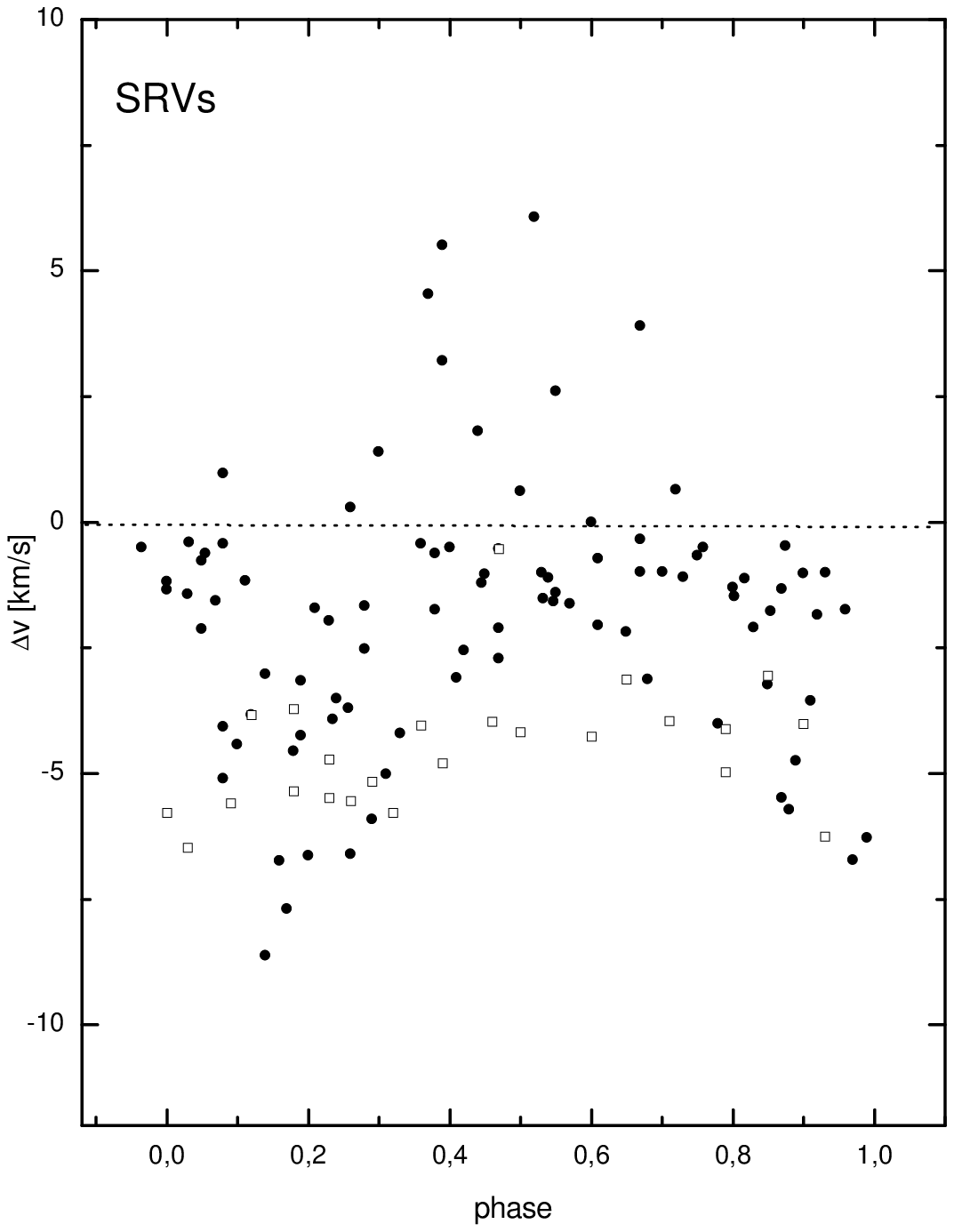}
\caption{Left: Velocity variations in Mira-type variables relative to the center
of mass velocity versus phase. Phases are derived from AAVSO lightcurves.  
Right: Velocity variations in semiregular variables relative to the center of
mass velocity versus phase. Phases are estimated from nearby light maxima or
minima observed with an automatic telescope. Open symbols denote the time series
of observations of W\,Cyg indicating an extreme case for the asymmetry relative
to the center of mass velocity.}
\end{center}
\end{figure}

The deep atmospheric layers of miras monitored by e.g.\,high excitation CO lines
at 1.6\,$\mu$m show a periodic variation with typical indications of radial
pulsation and the occurence of shocks (discontinuous velocity curves).  They
indicate the pulsation of the star that is also visible in the visual light
change. With the help of our large sample we are now able to derive mean
velocity curves for these layers in LPVs.  All miras (Fig.\,1, left panel) show
the same type of velocity variations. Velocity amplitudes are very similar. Our
sample of miras covers a period range between 145 and 470 days, and includes
stars of spectral type M, S and C.  No dependency on period, metallicity or
chemistry was found (Hinkle et al., 1997; Lebzelter et al.,
1999).

Semiregular variables (SRVs) have continuous velocity curves with the maximum
velocity occuring at light minimum (Fig.\,1, right panel). The velocity changes
show the same type of semiregular behaviour found in the light curves of these
stars.  Amplitudes range from 2 to 15 kms$^{-1}$ (Hinkle et al.\,1997; Lebzelter
1999).  A strong asymmetry relative to the center of mass velocity is observed
in most of the SRVs which is not understood yet (Lebzelter, 1999, Lebzelter
et al., 2000). One possible explanation, long term variations due to
pulsation or binarity, is discussed by Lebzelter et al.\ (these proceedings).

\begin{figure}
\begin{center}
\plotone{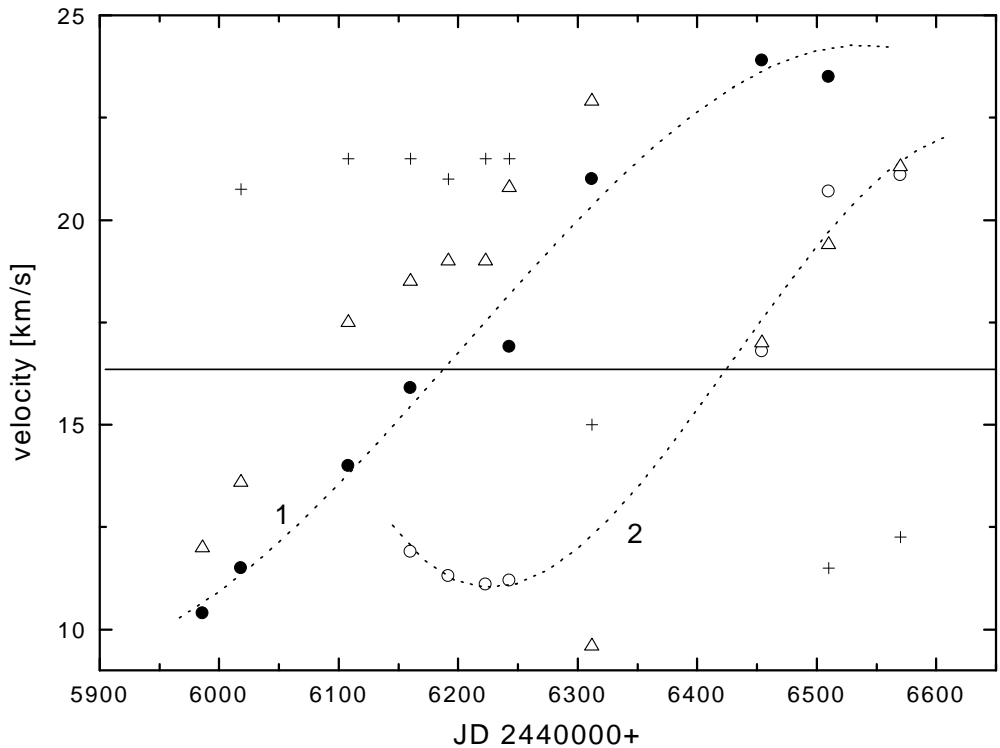}
\caption{H$_2$O and OH velocities of R\,Cas versus time. Filled and open circles
mark two different H$_2$O absorption components, crosses indicate water emission
lines, open triangles denote OH absorption. Two velocity curves have been drawn
through the H$_2$O data points. The horizontal line marks the centers of mass
velocity.}
\end{center}
\end{figure}

The outer atmospheric layers of these stars, investigated with the help of
molecules like SiO or H$_2$O, do not always follow the period defined by the
visual light change, i.e. line profiles obtained at similar phases can differ
significantly from each other.  These variations may be nonperiodic or periodic
on a longer timescale than the main period. An example is given in Fig.\,2 for
the velocities of high excitation H$_2$O lines in R Cas plotted against time. It
can be seen that the velocity curve starting at the beginning of the observed
time series shows the same shape as the typical mira velocity curves, but lasts
considerably longer than the light cycle. The next cycle shows again a velocity
curve following the stellar period. The observed difference between the two
cycles indicates that either the accelaration of the H$_2$O line forming layer
was higher or the deceleration lower. Both possibilities may be connected to a
non-continuous mass loss in this star (Lebzelter et al., 2001).  A similar
behaviour was found in a second star, $\chi$ Cyg. In both cases the outstanding
velocity behaviour followed a very bright light maximum. This suggests that
these two phenomena are related to each other. Richter \& Wood (2001) showed a
similar connection between bright maxima and outstanding behaviour of the UV
spectrum which is, as the 4\,$\mu$m lines, formed in the outer parts of the
atmosphere. However, this phenomenon seems to affect only the outer layers of
the star as such a behaviour is not observed e.g.\,in the 1.6\,$\mu$m CO
lines. It is therefore not due to a stronger pulsation of the star.

\section{Conclusions and outlook}

Periodic and nonperiodic velocity variations are found in AGB variables. The
inner layers close to the pulsation driving zone show a periodic and highly
similar velocity curve in all types of miras. This fact makes it very unlikely
that some of the miras pulsate in a different pulsation mode than the others as
has been discussed in the literature (e.g.\,Barthes, 1998). In the outer layers
nonperiodic effects occur and the observed stars show a strong diversity in
their line profile changes.

Our observational results allow a first overview on the pulsational properties
of the whole atmosphere of an AGB star. This can now be compared with recent
dynamic model atmospheres (H\"ofner, 1999).  Both periodic and nonperiodic
phenomena are predicted by the models. A detailled comparison of models and
observations has been started.

\end{document}